\documentclass[letterpaper,twocolumn,aps,superscriptaddress,english,showpacs,nofootinbib]{revtex4-2}
\usepackage[T1]{fontenc}
\usepackage[utf8]{inputenc}
\usepackage{geometry}
\usepackage{mathtools}
\usepackage{xcolor}
\usepackage{babel}
\usepackage{units}
\usepackage{braket}
\usepackage{amsmath}
\usepackage{amssymb}
\usepackage{stmaryrd}
\usepackage{graphicx}
\usepackage{wasysym}
\usepackage{bbold}
\usepackage{float}
\usepackage[colorlinks=true,citecolor=green,linkcolor =blue]{hyperref}
\usepackage[capitalise]{cleveref}
\usepackage[all]{hypcap} 
\usepackage{times}
\usepackage{orcidlink}

\makeatletter

\IfFileExists{lmodern.sty}{\usepackage{lmodern}}{}
\usepackage{hyperref}

\makeatother

\begin{document}
\title{
Quantum simulation of bubble nucleation across a quantum phase transition}

\author{De~Luo$^*$~\orcidlink{0000-0002-4842-0990}}
\affiliation{Duke Quantum Center, Department of Physics, and Department of Electrical and Computer Engineering, Duke University, Durham, NC 27701 USA}
\author{Federica~Maria~Surace$^*$~\orcidlink{0000-0002-1545-5230}}
\affiliation{Department of Physics and Institute for Quantum Information and Matter,
California Institute of Technology, Pasadena, California 91125, USA}
\author{Arinjoy~De}
\affiliation{Duke Quantum Center, Department of Physics, and Department of Electrical and Computer Engineering, Duke University, Durham, NC 27701 USA}
\author{Alessio~Lerose~\orcidlink{0000-0003-1555-5327}}
\affiliation{Rudolf Peierls Centre for Theoretical Physics, Clarendon Laboratory, Oxford OX1 3PU, United Kingdom}
\affiliation{Institute for Theoretical Physics, KU Leuven, Celestijnenlaan 200D, 3001 Leuven, Belgium}
\author{Elizabeth~R.~Bennewitz}
\affiliation{Joint Center for Quantum Information and Computer Science, NIST/University of Maryland, College Park, MD 20742 USA}
\affiliation{Joint Quantum Institute, NIST/University of Maryland, College Park, MD 20742 USA}
\author{Brayden~Ware}
\affiliation{Joint Center for Quantum Information and Computer Science, NIST/University of Maryland, College Park, MD 20742 USA}
\affiliation{Joint Quantum Institute, NIST/University of Maryland, College Park, MD 20742 USA}
\author{Alexander~Schuckert}
\affiliation{Joint Center for Quantum Information and Computer Science, NIST/University of Maryland, College Park, MD 20742 USA}
\affiliation{Joint Quantum Institute, NIST/University of Maryland, College Park, MD 20742 USA}
\author{Zohreh~Davoudi~\orcidlink{0000-0002-7288-2810}}
\affiliation{Joint Center for Quantum Information and Computer Science, NIST/University of Maryland, College Park, MD 20742 USA}
\affiliation{Maryland Center for Fundamental Physics and Department of Physics, University of Maryland, College Park, MD 20742, USA}
\author{Alexey~V.~Gorshkov}
\affiliation{Joint Center for Quantum Information and Computer Science, NIST/University of Maryland, College Park, MD 20742 USA}
\affiliation{Joint Quantum Institute, NIST/University of Maryland, College Park, MD 20742 USA}
\author{Or~Katz~\orcidlink{0000-0001-7634-1993}}
\affiliation{Duke Quantum Center, Department of Physics, and Department of Electrical and Computer Engineering, Duke University, Durham, NC 27701 USA}
\affiliation{School of Applied and Engineering Physics, Cornell University, Ithaca, NY 14853.}
\author{Christopher~Monroe \orcidlink{0000-0003-0551-3713}}
\affiliation{Duke Quantum Center, Department of Physics, and Department of Electrical and Computer Engineering, Duke University, Durham, NC 27701 USA}
\def\thefootnote{*}\footnotetext{These authors contributed equally to this work. \\Email: de.luo@duke.edu, fsurace@caltech.edu}
\def\thefootnote{$\dagger$}
\date{\today}

\begin{abstract}
The liquid-vapor transition is a classic example of a discontinuous (first-order) phase transition. Such transitions underlie many phenomena in cosmology, nuclear and particle physics, and condensed-matter physics. They give rise to long-lived metastable states, whose decay can be driven by either thermal or quantum fluctuations. Yet, direct experimental observations of how these states collapse into a stable phase remain elusive in the quantum regime. Here, we use a trapped-ion quantum simulator to observe the real-time dynamics of ``bubble nucleation'' induced by quantum fluctuations. Bubbles are localized domains of the stable phase which spontaneously form, or nucleate, and expand as the system is driven across a discontinuous quantum phase transition. Implementing a mixed-field Ising spin model with tunable and time-dependent interactions, we track the microscopic evolution of the metastable state as the Hamiltonian parameters are varied in time with various speeds, bringing the system out of equilibrium. Site-resolved measurements reveal the emergence and evolution of finite-size quantum bubbles, providing direct insight into the mechanism by which the metastable phase decays. We also identify nonequilibrium scaling behavior near the transition, consistent with a generalized Kibble-Zurek mechanism. Our results demonstrate the power of quantum simulators to probe out-of-equilibrium many-body physics, including quantum bubble nucleation, a key feature of discontinuous quantum phase transitions, with application to studies of matter formation in the early universe.
\end{abstract}

\maketitle

The concept of metastability is connected to some of the most profound questions in physics, including the stability of our universe itself~\cite{turner1982our}. For example, the electroweak vacuum is conjectured to be metastable~\cite{coleman,callan1977fate}. This possibility is explored through the theory of false-vacuum decay~\cite{coleman,callan1977fate}, which describes transitions from a metastable state to a more stable one. Understanding metastable decay, which occurs during discontinuous (first-order) phase transitions, could provide insights into early-universe physics and cosmic inflation~\cite{Kibble1980, Guth1981}, and the origin of baryogenesis~\cite{cohen1991baryogenesis,anderson1992electroweak,Baryogenesis}, primordial black holes~\cite{liu2022primordial,davoudiasl2022supermassive}, and dark matter~\cite{baker2020filtered,hong2020fermi}, with possible gravitational-wave signals~\cite{hindmarsh2014gravitational,weir2018gravitational}. Metastability also arises in condensed matter physics, including supercooled liquids \cite{debenedetti1996metastable,CAVAGNA200951,Debenedetti2001,Berthier2011}, crystal melting \cite{wunderlich1973macromolecular}, and frustrated spin models \cite{Villain1979,Lacroix2011,Moessner2006}. 

Transition from a metastable state to a more stable state occurs through ``bubble nucleation''~\cite{langer, Oxtoby_1992}. This process underlies key phenomena, such as string breaking in confining theories~\cite{Surace2024} and the Schwinger process of particle-antiparticle pair production~\cite{Schwinger1951,coleman1976more}. For decays driven by quantum fluctuations, the central mechanism governing these processes involves tunneling through an energy barrier between the metastable state and the true ground state. Because overcoming such energy barriers requires changes over large spatial regions, metastable states can persist out of equilibrium for long times \cite{Lerose2020,Chao2023,yin2025theorymetastablestatesmanybody}.

Studying nonequilibrium dynamics near discontinuous quantum phase transitions (QPTs) can be computationally challenging due to the long time scales involved \cite{Sinha2021,lagnese2021,Lagnese2024,Darbha2024}. Quantum simulators provide precise control over system parameters, enabling the study of complex quantum phenomena both in and out of equilibrium. In particular, they provide powerful tools for investigating metastable phenomena in nonequilibrium dynamics \cite{Ng2021,Schuckert2023,Darbha2024,zhu2024probingfalsevacuumdecay,vodeb2024stirring}. Quantum gas experiments have already probed discontinuous QPTs in continuous space, demonstrating key features such as metastability, false vacuum decay, and dynamical scaling across the transition~\cite{Qiu2020,Song2022,zenesini2024false}. However, the control necessary to study the coherent formation of bubbles with spatial and temporal resolution has not been demonstrated so far. In particular, previous observations of bubble nucleation could not reach the low temperatures required to probe the regime dominated by quantum fluctuations \cite{zenesini2024false}; probe large, metastable bubbles~\cite{zhu2024probingfalsevacuumdecay}; or achieve sufficiently low dephasing to maintain coherent evolution for long times~\cite{vodeb2024stirring}.

In this work, we use a trapped-ion quantum simulator to investigate, for the first time, coherent quantum dynamics across a discontinuous quantum phase transition with spatio-temporal resolution. As an example, we consider the phenomenon of string breaking, recently explored both theoretically \cite{Verdel19_ResonantSB,Verdel2023,Surace2024,mallick2024stringbreakingdynamicsising,borla2025stringbreaking21dmathbbz2} and experimentally \cite{De2024,cochran2024,gonzalezcuadra2024,Ciavarella2024,liu2024string,crippa2024analysis, alexandrou2025realizing} in quantum spin systems and lattice gauge theories. In quantum chromodynamics, string breaking occurs when a flux tube, i.e., string, connecting two static probe quarks becomes unstable. The string decays via the production of additional quarks that bind to the probe quarks. We simulate a simplified one-dimensional model of this process. Using our simulator's spatio-temporal control, we prepare a string and gradually increase the string tension, driving it into an unstable state. This triggers the string to break, a process modeled as the crossing of a discontinuous QPT under specific boundary conditions~\cite{Surace2024,ROSSINI20211}. We study the resulting nonequilibrium dynamics, in particular searching for signs of bubble formation and scaling laws.
\begin{figure*}[t]
    \includegraphics{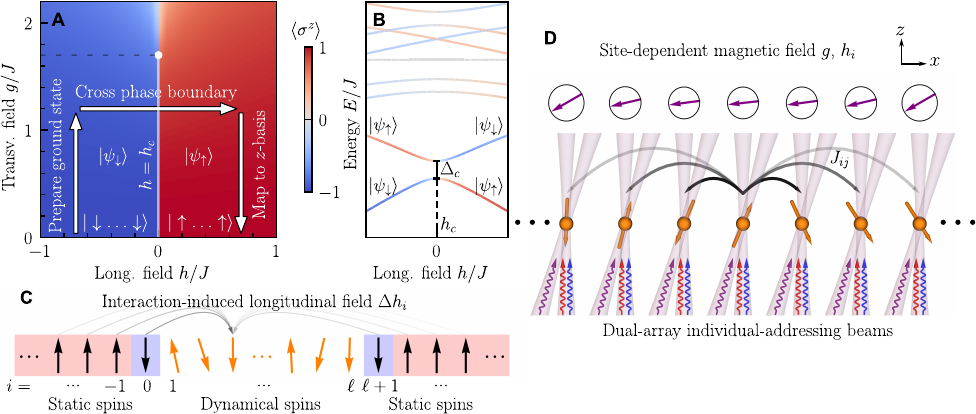}
    \caption{\textbf{Phase diagram of the mixed-field Ising model and implementation with trapped ions.} (\textbf{A}) Ground-state phase diagram of the Hamiltonian in Eq.~(\ref{eq_H}) as a function of the transverse field $g$ and longitudinal field $h_i=h$ for  $\beta=1.21$ in the thermodynamic limit. The color scale indicates the order parameter $\langle\sigma^z\rangle$ defined in Eq.~(\ref{eq:orderpar}). The vertical line indicates the discontinuous phase transition ($h=h_c=0$ and $g<g_*$) and the horizontal dashed line at $g=g_*=1.7J$ ends at the continuous phase-transition point at $h=h_c$. The arrows indicate the ramps used in the quantum simulation. (\textbf{B}) Lowest energy levels of the Hamiltonian in Eq.~(\ref{eq_H}) along the $h$-ramp for $\beta=1.21$, $g=1.2J$, and $\ell=5$. The lines are colored by the order parameter $\langle \sigma^{z}\rangle$ using the same color scale as in \textbf{A}. The $\ket{\psi_\downarrow}$ and $\ket{\psi_\uparrow}$ eigenstates form an avoided crossing with gap $\Delta_c$ at the discontinuous quantum phase transition at $h_c$. (\textbf{C}) Ising spin chain and boundary conditions in the quantum simulation. The dynamical spins shown in orange are simulated by the quantum simulator. Two domain walls are pinned adjacent to the dynamical spins. Interactions with the static spins, indicated by the gray arrows, are emulated by a site-dependent longitudinal field acting on the dynamical spins, which is individually controlled over time. (\textbf{D}) Experimental realization of the Hamiltonian. Purple arrows indicate the orientation and strength of the site-dependent mixed magnetic field. Gray arrows indicate (a set of) all-to-all spin-spin interactions. Wiggly arrows in the individual-addressing beams represent the beatnote frequencies generated by the beams, which drive the interaction and magnetic-field terms in the Hamiltonian (see Methods, Sec.~\ref{sec_exp}).}
    \label{fig_intro}
\end{figure*}

\emph{Model and experimental setup.}---We study a paradigmatic model widely used in studies of quantum phase transitions \cite{Sachdev2011}, a mixed-field Ising chain of $\ell$ spins described by the Hamiltonian
\begin{equation}\label{eq_H}
H=-\sum_{i<j}^{\ell}J_{i,j}\sigma^{z}_{i}\sigma^{z}_{j}-\sum_{i=1}^{\ell}h_i\sigma^{z}_{i}-g\sum_{i=1}^{\ell} \sigma^{x}_{i},
\end{equation}
where $\sigma^\alpha_i$ are Pauli matrices on spin $i$, $h_i$ and $g$ represent the longitudinal- and transverse-field components, respectively, and $J_{i,j} \coloneq J\exp{[-\beta (|j-i|-1)]}$ are exponentially decaying ferromagnetic couplings with decay constant $\beta$. Here, $J$, $\beta$, and $g$ are positive. The ground-state phase diagram of this model for a uniform longitudinal field $h_i=h$ is presented in Fig.~\ref{fig_intro}A. In the absence of a longitudinal field ($h=0$), the model has a global $\mathbb{Z}_2$ symmetry generated by $\prod_{j=1}^\ell\sigma_j^x$ and features a continuous QPT at a critical value $g=g_*$ (see Methods, Sec.~\ref{subsec:phasetrans}). Explicitly, the order parameter 
\begin{equation}
\label{eq:orderpar}
\langle\sigma^{z}\rangle \coloneq \tfrac{1}{\ell}\sum_{i=1}^{\ell}\langle\sigma^{z}_i\rangle
\end{equation}
changes continuously from the ferromagnetic phase at $g<g_*$ where $\langle\sigma^{z}\rangle\neq0$, to a paramagnetic phase at $g>g_*$ where $\langle\sigma^{z}\rangle=0$. 

Introducing a nonzero longitudinal field $h$ breaks the energy degeneracy in the ferromagnetic phase, resulting in a discontinuous QPT between the ground state with negative magnetization $\left|\psi_{\downarrow}\right\rangle$ and the ground state with positive magnetization $\left|\psi_{\uparrow}\right\rangle$ as $h$ is varied across the critical line $h_c=0$. In the limit of $|g|\ll J$, these two states become the fully polarized spin states $\left|\downarrow\dots\downarrow\right\rangle$ and $\left|\uparrow\dots\uparrow\right\rangle$ in the $z$ basis.\footnote{We refrain from carrying a $z$ subscript to denote the states' chosen basis, since all states are assumed in the $z$ basis throughout, with one exception: In Methods, Sec.~\ref{sec_exp}, we restore such basis subscripts to distinguish the conventional experimental computational basis from that used in the main paper.} Around $h=h_c$, the order parameter  $\langle\sigma^{z}\rangle$ changes sign very rapidly with $h$ for finite $\ell$, and the transition becomes discontinuous in the thermodynamic limit $\ell\rightarrow\infty$. The energy spectrum for $\ell=5$ is shown in Fig.~\ref{fig_intro}B. In a finite system, the QPT arises from an avoided crossing at $h_c$ between the two lowest energy levels, corresponding to the states $\left|\psi_{\downarrow}\right\rangle$ and $\left|\psi_{\uparrow}\right\rangle$. The minimum energy gap $\Delta_c$ is proportional to the matrix element $\langle \psi_\uparrow|H|\psi_\downarrow\rangle$, which represents the transition amplitude corresponding to flipping all the spins. As shown in Methods, Sec.~\ref{subsec:phasetrans} and Fig.~\ref{fig_finite_size}, $\Delta_c\propto(g/g_*)^\ell$ \cite{Surace2024}. To study the quantum many-body dynamics of this model, we focus on a regime where $g\approx J$ (away from the classical limit $g\ll h,J$ where quantum fluctuations are negligible). 

While the phase diagram in Fig.~\ref{fig_intro}A captures the general equilibrium properties of the mixed-field Ising model, additional insights can be gained by introducing boundary conditions that link the discontinuous QPT to string breaking (See Methods, Sec.~\ref{sec_Z2}). This link relies on the interpretation of the Ising model as a lattice gauge theory, where a domain wall (i.e., the boundary between two anti-aligned adjacent spins) acts as a charged particle (or ``quark,'' in analogy with quantum chromodynamics) and a domain of $\downarrow$ spins plays the role of an electric-field flux, i.e.~a string, connecting the charges. In this analogy, $J$ corresponds to particle mass, $h$ to string tension, and $g$ to coupling strength~\cite{De2024}. The string-breaking setup enables the study of the spatial structure of the time-evolved state, distinguishing edge and bulk features associated with the decay of a metastable string.

We consider boundary conditions that emulate two static domain walls pinned at the ends of the chain \cite{De2024,Surace2024}, as shown in Fig.~\ref{fig_intro}C. For $h<h_c$, this setup induces a flux string connecting the two static quarks, corresponding to the eigenstate $\ket{\psi_\downarrow}$. If the string tension $h$ exceeds the critical threshold $h_c$, string breaking becomes energetically favorable, creating two additional domain walls at the edges, generating state $\ket{\psi_\uparrow}$. This formulation provides a controlled environment for studying the evolution of the string and its breaking in real time~\cite{De2024}. Here, we aim to study metastable states emerging in such dynamics.

The fictitious semi-infinite chains of static spins induce a site-dependent longitudinal field on the dynamical spins:
\begin{equation}
\label{eq:dh}
\Delta h_i=\big(\sum_{j=-\infty}^0+\sum_{j=\ell+1}^\infty\big)J_{i, j}\langle\sigma_{j}^{z}\rangle,~~~ 1 \leq i \leq \ell.
\end{equation}
This field decays exponentially from the boundaries and explicitly breaks the $\mathbb{Z}_2$ symmetry (see Methods, Sec.~\ref{sec_Z2}). The total longitudinal field acting on the $i$-th spin among the $\ell$ physical spins thus reads $h_i=h+\Delta h_i$. As a result, the transition point for a finite system, i.e., the point with the smallest spectral gap, shifts to $h_c(\ell)> 0$ for chains of finite size $\ell$, recovering $h_c=0$ only in the limit $\ell\rightarrow \infty$.

We experimentally probe this transition and its dynamics using a trapped-ion quantum simulator, as illustrated in Fig.~\ref{fig_intro}D. Our setup consists of a one-dimensional crystal of $^{171}$Yb$^{+}$ ions in a linear surface trap, where spin states are encoded in the electronic ground-state clock levels of the ions. A global laser beam and a dual array of tightly focused beams form Raman beatnotes that generate the Hamiltonian terms, as shown by the wiggly arrows in Fig.~\ref{fig_intro}D. The tones shown in red and blue, detuned from the motional sideband transitions, generate the programmable Ising interactions with tunable $\beta$. The tone shown in purple, on resonance with the qubit transition, generate the longitudinal fields. A common detuning on all three tones generates the tranverse fields (see Methods, Sec.~\ref{sec_exp}). A key experimental advance in this work is the simultaneous and fully programmable control of both transverse and longitudinal fields at the level of individual sites, with the ability to dynamically modulate their amplitudes in time. This capability enables the implementation of spatially inhomogeneous Hamiltonians, used for the study of nonequilibrium quantum dynamics across a discontinuous QPT, in a way previously inaccessible.

\emph{Locating the transition point.}---As mentioned above, the boundary conditions with two static domain walls [inducing the site-dependent longitudinal field given in Eq.~(\ref{eq:dh})] result in a shift in the transition point $h_c(\ell)$ for finite $\ell$. We study two complementary methods to experimentally identify the transition point.

\begin{figure}[t!]
    \includegraphics[width=\linewidth]{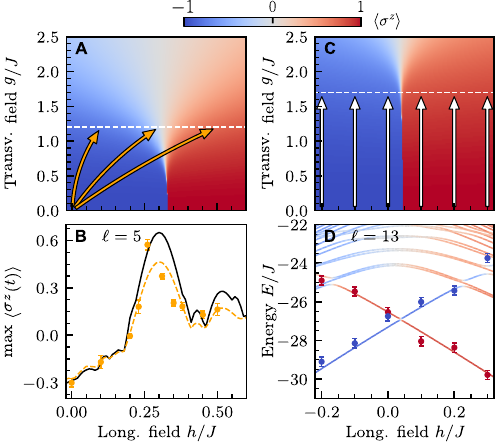}    \caption{\textbf{Measurement of phase boundary $\boldsymbol{h_c}$.} Boundary conditions emulate two static domain walls as in Fig.~\ref{fig_intro}C. (\textbf{A}) For $\ell=5$ and $\beta=1.21$, we quench the Hamiltonian from $h=g=0$ to $0\leq h\leq 0.5J$ and $g=1.2J$, indicated by the orange arrows, and measure the dynamics after the quench (see Fig.~\ref{fig_quench_data}). (\textbf{B}) For the setting described in \textbf{A}, the maximum value of $\langle \sigma^z \rangle$ vs.~$h$ shows a peak at $h_c$. The solid black (dashed orange) curve corresponds to numerical simulation without (with) decoherence (see Methods, Sec.~\ref{sec_sim}). (\textbf{C}) For $\ell=13$ and $\beta=0.78$, we prepare the eigenstates $\ket{\psi_\downarrow}$ and $\ket{\psi_\uparrow}$ by ramping $g$ from $0$ to $1.7J$, indicated by the white arrows, with the initial states $\ket{\downarrow\dots\downarrow}$ and $\ket{\uparrow\dots\uparrow}$, respectively. (\textbf{D}) For the setting described in \textbf{C}, the measured energies of $\ket{\psi_\downarrow}$ (blue) and $\ket{\psi_\uparrow}$ (red) as a function of $h$, are overlaid with the eigenenergy spectrum. A constant offset is added to the measured values to account for single-qubit errors (see Fig.~\ref{fig_13ion_energy}). Points in \textbf{B} and \textbf{D} correspond to experimental data with 250 and 2000 repetitions, respectively, and the error bars indicate standard deviation from $N=1000$ bootstrap samples.}
    \label{fig_hc}
\end{figure}

The first technique employs a quench, i.e., a sudden change in Hamiltonian parameters. This method allows us to probe the resonant transitions between $\ket{\psi_\downarrow}$ and $\ket{\psi_\uparrow}$ at $h_c$, and is straightforward to implement in experiment. Specifically, we initialize the system in state $\ket{\downarrow\dots \downarrow}$, corresponding to the ground state at $h=g=0$, and suddenly quench the Hamiltonian in Eq.~(\ref{eq_H}) to a fixed $g>0$ and a variable $h>0$, as indicated by the arrows in Fig.~\ref{fig_hc}A for an $\ell=5$ chain and $\beta=1.21$.

The response of the system after the quench depends on the overlap between the initial state and the energy eigenstates of the final Hamiltonian. As shown in Fig.~\ref{fig_quench_data}, away from the transition point, the initial state predominantly overlaps with $\ket{\psi_\downarrow}$, which is an eigenstate, hence small amplitude oscillations in $\langle\sigma^{z}(t)\rangle$. However, at the avoided crossing $h\approx h_c$, the initial state overlaps equally with the two lowest-energy eigenstates, which are superpositions of $\ket{\psi_\uparrow}$ and $\ket{\psi_\downarrow}$. This results in coherent oscillations of $\langle\sigma^{z}(t)\rangle$ at frequency $\Delta_c$, reflecting interference between the two magnetization patterns, see Methods, Sec.~\ref{sec_finite_size}.
In Fig.~\ref{fig_hc}B, we extract the maximal observed value of $\langle\sigma^{z}\rangle$ from each quench experiment and plot it as a function of $h$. The resonance feature, which appears as a peak in this plot, clearly identifies the transition point at $h_c = 0.31J$, consistent with numerical simulations.

For larger system sizes, the energy gap $\Delta_c$ decreases exponentially, diminishing the amplitude and frequency of oscillations near the gap (see Methods, Sec.~\ref{sec_finite_size}), making these oscillations impractical to resolve. To estimate $h_c$ for longer chains, we demonstrate a second technique that applies ramps of the transverse field $g$, as illustrated in Fig.~\ref{fig_hc}C, to prepare the states $\ket{\psi_{\downarrow}}$ and $\ket{\psi_{\uparrow}}$, and measures their energy. For each value of $h$, we initialize the system in each of the two fully polarized states, which are eigenstates at $g=0$. We then linearly ramp $g$ from $0$ to $1.7J$ and measure the energy of the resulting state. We verify that the system adiabatically follows the instantaneous eigenstate during the ramp by comparing the measured energy with the numerically evaluated eigenenergy.\footnote{We verify that, unlike the classical state (i.e., $g=0$), the prepared states exhibit nonzero connected spin-spin correlations $C_{i,j}=\langle\sigma^z_i\sigma^z_j\rangle-\langle\sigma^z_i\rangle\langle\sigma^z_j\rangle$, indicating the presence of quantum correlations (see Fig.~\ref{fig_ground_state_corr}).} In Fig.~\ref{fig_hc}D, we present the measured energies for $\ell=13$, $\beta=0.78$, and a linear ramp with time $T=2.4/J$ up to a value $g=1.7 J$. The measured energies are overlaid on the calculated spectrum of this system, taking into account single spin-flip errors (see Fig.~\ref{fig_13ion_energy}). 
The crossing of the two energy curves thus provides $h_\textrm{c}$. Although this method is more experimentally demanding than the quench technique, it is less restrictive, as the required evolution time does not depend on $\Delta_c$.

\emph{Dynamics across the transition}.---The technique in the previous section allowed efficient preparation of the phases for long chains and enabled precise estimation of the string-breaking point, where the ground state changes abruptly. We now study the nonequilibrium dynamics of the system as the field $h$ is ramped across the phase boundary. Since both the energy gap $\Delta_c$ and the adiabaticity of the process, i.e., the $h$-ramp duration $\tau_h$, depend strongly on system size $\ell$, we expect two qualitatively different behaviors. For small systems, the evolution may be more or less adiabatic depending on the ramp speed. If the ramp is slow enough, the system remains in its ground state, flipping its magnetization at the transition and settling into the ground state at larger $h$. In this regime, further increasing $h$ does not induce additional transitions. For larger systems, maintaining adiabatic evolution becomes impractical. The system remains in $\ket{\psi_{\downarrow}}$ past $h_c$, but enters an unstable state that undergoes multiple level crossings as $h$ further increases. This leads to excitations into higher-energy states (see Methods, Sec.~\ref{sec_spectrum}). This evolution can be interpreted as the nucleation of ``bubbles'' (i.e. domains with positive $\langle \sigma^z \rangle$) of various sizes that progressively form as $h$ increases. The dynamical string-breaking process thus occurs through local breakings, rather than the global breaking characteristic of the ground-state transition.
\begin{figure}[t!]
    \includegraphics[width=\linewidth]{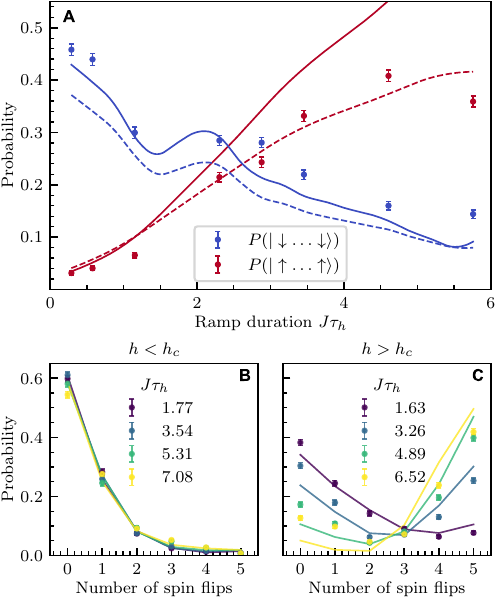}
    \caption{\textbf{ 
    Probe of the ground-state phase transition for $\boldsymbol{\ell=5}$ and $\boldsymbol{\beta=1.21}$.} 
    In all subplots, points correspond to experimental data with 2000 repetitions, and the error bars indicate the standard deviation from $N=1000$ bootstrap samples. 
    (\textbf{A}) We use the ramp procedure shown in Fig.~\ref{fig_intro}A, with $g=1.2J$ and $h=1.0J$. The $h$-ramp trajectory is optimized, as shown in Fig.~\ref{fig_ramp}, and the full $h$-ramp duration $\tau_h$ is varied. Red (blue) color indicates $P_\uparrow \equiv P(\ket{\uparrow\dots\uparrow})$ ($P_\downarrow \equiv P(\ket{\downarrow\dots\downarrow})$), i.e., the probability of all spins (no spins) flipped starting from the $\ket{\downarrow\dots\downarrow}$ initial state. 
    Solid (dashed) lines correspond to numerical simulations without (with) decoherence (see Methods, Sec.~\ref{sec_sim}).
    (\textbf{B-C}) The probability for given numbers of flipped spins, starting from the initial state $\ket{\downarrow \dots \downarrow }$, measured in a state with the $h$-ramp terminated at $h=0.2 J<h_c$ (\textbf{B}) and at $h=0.5J>h_c$ (\textbf{C}). Here, $g$ is adiabatically ramped up to $g=1.2J$ but is not ramped down to zero. Colors correspond to different full $h$-ramp duration $\tau_h$ from $h=0$ to $h=1.0J$. For \textbf{B} and \textbf{C}, simulation results with decoherence are not shown because they nearly overlap with the decoherence-free results shown in solid lines.
    }
    \label{fig_ramp_5ions}
\end{figure}

These transition dynamics can be studied using a three-stage protocol illustrated in Fig.~\ref{fig_intro}A. With the initial state $\ket{\downarrow\dots\downarrow}$ at $h=0$, we adiabatically ramp $g$ from $0$ to a positive value in the first stage to prepare the eigenstate $\ket{\psi_\downarrow}$. Then we ramp $h$ to a final value $h>h_c$ in the second stage and vary the ramp duration $\tau_h$. Lastly, we adiabatically ramp $g$ back to zero and measure the state in the $z$ basis. The temporal shapes of the $g$- and $h$-ramps are engineered, by varying the ramp rate as a function of the instantaneous gap, to enhance adiabaticity  (see Methods, Sec.~\ref{sec_ramp}). While the evolution during the $g$-ramps is approximately adiabatic, the small gap at $h=h_c$ hinders adiabatic evolution during the $h$-ramp. 

For small chains and slow ramps, dynamics primarily involve transitions between $\left|\psi_{\downarrow}\right\rangle$ and $\left|\psi_{\uparrow}\right\rangle$, and can be approximated by a Landau-Zener process.  The populations of these two states after the $h$-ramp can be measured upon ramping $g$ back to 0, such that $\left|\psi_{\downarrow}\right\rangle$ and $\left|\psi_{\uparrow}\right\rangle$ evolve adiabatically into $\ket{\downarrow\dots\downarrow}$ and $\ket{\uparrow\dots\uparrow}$.
In Fig.~\ref{fig_ramp_5ions}A, we probe the Landau-Zener behavior by measuring the occurrence probabilities $P_\downarrow\equiv P(\ket{\downarrow\dots\downarrow})$ and $P_\uparrow\equiv P(\ket{\uparrow\dots\uparrow})$ as functions of $\tau_h$ for an $\ell=5$ chain with $\beta=1.21$. Here, $h_c=0.31J$, and $h$ is ramped from $0$ to $1.0J$ at $g=1.2J$. For fast ramps, the population remains mostly in $\ket{\psi_\downarrow}$, while for slow ramps, it follows the ground state, resulting in collective spin flipping. As $\tau_h$ increases, $P_\uparrow$ gradually overtakes $P_\downarrow$, and the system transitions from the string phase to the broken-string phase. At longer evolution times, decoherence effects become significant and limit the feasibility of experiments with larger $\tau_h$ values.

To further characterize the crossing of the transition, we compare the distribution of the number of flipped spins before and after crossing $h_c$, as shown in Fig.~\ref{fig_ramp_5ions}B-C. Here, we ramp $g$ from $0$ to $1.2J$ and measure the $z$-basis state distributions at snapshots along the $h$-ramp (points along the horizontal path in Fig.~\ref{fig_intro}A) without the final $g$-ramp. At points $h<h_c$ ($h>h_c$) below (above) the critical threshold, we compare the distribution of spin flips for various $h$-ramp durations $\tau_h$, where $\tau_h$ is defined for the full $h$-ramp from $h=0$ to $h=1J$. For $h=0.2J<h_c$, the state histogram is nearly independent of $\tau_h$, whereas for $h=0.5J>h_c$, the distribution varies significantly with ramp duration, reflecting increased spin flips for slower ramps. (Note that the population in $\ket{\downarrow\dots\downarrow}$ and $\ket{\uparrow\dots\uparrow}$ is not equivalent to population in $\ket{\psi_\downarrow}$ and $\ket{\psi_\uparrow}$ at $g> 0$.)
This comparison illustrates that the small gap at the avoided crossing places stringent constraints on the ramp speed required for adiabatic passage through the transition.

\emph{Bubble nucleation.}---For larger system sizes, adiabatic evolution across the phase boundary is suppressed by the exponentially small energy gap $\Delta_c$. Repeating the same measurement for $\ell=13$ and applying the same ramping durations as for the $\ell = 5$ system exhibits negligibly small probability of populating the ground state, as shown in Fig.~\ref{fig_13ion_ramp_adiabatic}. At the same time, the transition point $h_c(\ell)$ gets close to zero for longer chains. While the collective flip of all the $\ell$ spins at $h_c$ is highly suppressed, increasing $h$ beyond the critical value leads to successive level crossings, involving localized spin flips in smaller domains, as shown in Fig.~\ref{fig_N13Spectrum}.
\begin{figure}
    \includegraphics[width=\linewidth]{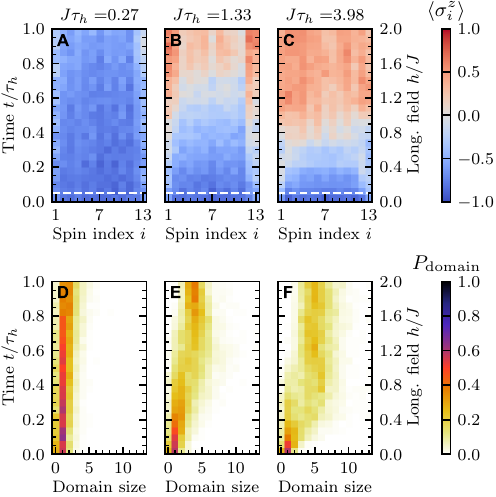}
    \caption{\textbf{Phase-transition dynamics for $\boldsymbol{\ell=13}$ and $\boldsymbol{\beta=1.21}$.} Starting with $\ket{\psi_\downarrow}$ at $g=1.2J$, we linearly ramp $h$ from $0$ to $2.0J$ and measure the dynamics during the ramp. Time evolution of individual-spins magnetization, $\langle \sigma_i^z \rangle$, is plotted in (\textbf{A-C}), and of the probability that the largest connected domain of $\uparrow$-spins has size $n$, $P_\text{domain}$, is plotted in (\textbf{D-F}), for $J\tau_h=0.27$, $1.33$, and $3.98$, respectively.
    Each row is an average of 500 experimental repetitions.
    The dashed white lines in \textbf{A-C} indicate the moment when $h=h_c$.}
    \label{fig_ramp_13ions}
\end{figure}

To characterize these excitations, we perform a linear ramp of $h$ according to $h(t)/J=2t/\tau_h$, and analyze the resulting spin-domain patterns. Following the same initial ramp of $g$ from 0 to $1.2J$ to prepare $\ket{\psi_\downarrow}$, we measure real-time spin dynamics in the $z$ basis along the $h$-ramp. We do not ramp $g$ back down, since higher-energy levels populated during the $h$-ramp cannot be unambiguously connected with the $g=0$ eigenstates due to multiple level crossings. Figures~\ref{fig_ramp_13ions}(A-C) shows the local magnetization $\langle \sigma_i^z \rangle$ as the longitudinal field $h$ is ramped from $0$ to $2J$, well beyond the transition point $h_c(\ell=13)=0.1J$. For short ramp durations, the magnetization remains nearly constant. For slower ramps, it changes sign at $h>h_c(\ell)$, indicating a transition via bubble formation---localized domains of positive magnetization---rather than a collective spin flip (see Methods, Sec.~\ref{sec_spectrum}). Additionally, magnetization changes at the edges prior to changing in the bulk, reflecting the influence of static quarks at the boundaries. This suggests that string breaking initiates preferentially near static boundary quarks~\cite{
De2024, Surace2024}.

To further characterize bubble nucleation, we analyze the growth of the largest domain size as a function of ramp time. Figures~\ref{fig_ramp_13ions}(D-F) display $P_\text{domain}(n)$, the probability that the largest connected domain of $\uparrow$-spins has size $n$, as a function of domain size $n$ and time elapsed during the $h$-ramp. For a very short ramp duration $J\tau_h=0.27$, bubble formation is minimal, and the largest domain size is typically 1. For longer ramp times $J\tau_h=1.33$ and $3.98$, domain sizes grow during the ramp, with $P(n)$ peaked around $n=4$ and $n=5$ at the end of the ramp, respectively. The result agrees well with numerical simulations, as shown in Fig.~\ref{fig:13ionsim}. The maximum domain size is constrained by the ramp's adiabaticity relative to the energy gaps at the level crossings, with larger domains forming for slower ramps.
\begin{figure}
    \includegraphics[width=\linewidth]{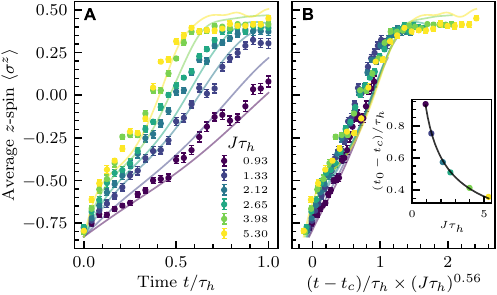}
    \caption{\textbf{Scaling law at the discontinuous quantum phase transition for $\boldsymbol{\ell=13}$, $\boldsymbol{g=1.2J}$, and $\boldsymbol{\beta=1.21}$}. (\textbf{A}) Dynamics of $\langle\sigma^z\rangle$ during the $h$-ramp for a range of $\tau_h$, using the same linear-ramp protocol as in Fig.~\ref{fig_ramp_13ions}. (\textbf{B}) Rescaled time traces from \textbf{A} collapse to a single curve using a power-law scaling with exponent $\mu=0.56$. Inset: The time difference $|t_0 - t_c|$ (where $t_0$ is when $\langle\sigma^z\rangle$ changes sign and $t_c$ is when $h(t) = h_c$) versus $\tau_h$ fits a power law with exponent $\mu$ (black line). In \textbf{A} and the main panel of \textbf{B}, points represent averages over 500 experimental repetitions, with error bars showing standard deviations from $N=1000$ bootstrap samples. Solid lines indicate numerical simulations without decoherence. Points in the inset in \textbf{B} show numerical simulation results.}
    \label{fig_scaling_law}
\end{figure}

\emph{Generalized Kibble-Zurek scaling.}---Crossing a continuous QPT often follows universal scaling laws governed by the Kibble-Zurek mechanism~\cite{Kibble1980, Zurek1985}. As a system is driven through a symmetry-breaking transition point at a finite rate, the energy gap closes. This causes the evolution to become nonadiabatic near the transition, and leads to the formation of domains in the symmetry-broken phase. Both the characteristic size of these domains and the timescale over which they form exhibit a power-law dependence on the rate of the drive. The Kibble-Zurek mechanism has been verified across multiple experimental platforms, including ultracold atoms~\cite{Anquez2016, Clark2016}, trapped ions~\cite{Li2023}, and Rydberg-atom arrays~\cite{Keesling2019}. However, its applicability to discontinuous quantum phase transitions, at least in the far-from-adiabatic regime~\cite{Pelissetto2020,Pelissetto2023,ROSSINI20211}, remains an open question. Generalized Kibble-Zurek scaling laws may describe dynamics near the transition point of certain discontinuous phase transitions~\cite{Qiu2020}.

We observe a power-law scaling for the ramp dynamics of $\ell=13$, as shown in Fig.~\ref{fig_scaling_law}, using the same linear-ramp protocol as in Fig.~\ref{fig_ramp_13ions}. For each ramp duration $\tau_h$, define the time $t_{0}$ as the moment during the ramp when $\langle \sigma^z\rangle$ changes sign, with the corresponding longitudinal-field value $h=h_0$; and $t_c$ as the moment when $h=h_c$. The time interval $|t_{0}-t_c|/\tau_h \equiv |h_0-h_c|/(2J)$ is well described by the function $\tau_h^{-\mu}$ and our fit gives $\mu=0.56$, as shown in the inset of Fig.~\ref{fig_scaling_law}B. Rescaling the time interval $|t-t_c|/\tau_h$ by $\tau_{h}^\mu$ collapses the time evolution of $\langle \sigma^z\rangle$ onto a single curve across a wide range of ramp durations. Further studies could elucidate whether this scaling behavior fits within a generalized Kibble-Zurek framework.

\emph{Conclusion.}---We have experimentally studied a discontinuous quantum phase transition in a mixed-field Ising spin model using a fully programmable trapped-ion quantum simulator. By mapping the phase boundary across different system sizes, we examined the system's nonequilibrium evolution across the transition and investigated transition's finite-size scaling. For small systems, where adiabatic evolution is possible, the dynamics occurs in a Landau-Zener regime, while for larger systems, the transition occurs via bubble-nucleation dynamics, consistent with a generalized Kibble-Zurek mechanism. The observed transition can also be interpreted as a string-breaking process, analogous to the decay of a flux string between a quark-antiquark pair. In previous work, we studied the post-quench dynamics of string breaking, focusing on how the system evolves after a sudden parameter change~\cite{De2024}. Here, we took a complementary approach, exploring controlled processes---including adiabatic and diabatic protocols---across the transition. This framework deepens our understanding of string-breaking dynamics, in particular revealing universal features in the form of a scaling law.

The tunability of interactions and system parameters in the platform of this work provides new opportunities to probe metastability, nonequilibrium dynamics, and quantum critical phenomena in strongly correlated systems, including understanding the origin and generality of the scaling law observed. The connection between the phenomena we studied and analogous situations in nuclear and high-energy physics, cosmology, and condensed-matter physics underscores the potential of trapped-ion quantum simulators to explore quantum phases and phase transitions in nature. Future experiments could extend these studies to larger system sizes, more complex quantum-field and gauge-theoretic settings, and real-time dynamics of models of early universe and high-energy particle collisions.

\renewcommand{\figurename}{Figure}
\renewcommand{\thefigure}{S\arabic{figure}}
\renewcommand{\theHfigure}{S\arabic{figure}}
\setcounter{figure}{0}

\section*{Methods}

\subsection{Finite-size scaling of the phase transition}\label{sec_finite_size}
\label{subsec:phasetrans}

This section describes how the location of the phase-transition point $h_c(\ell)$ can be determined for finite spin chains and how it depends on system size. 

To model the dynamics near $h_c$, one can define an effective two-level Hamiltonian for the lowest-energy states of the form
\begin{align}
    H_{\mathrm{eff}}=&M(h-h_c)\left(\ket{\psi_\downarrow}\bra{\psi_{\downarrow}}-\ket{\psi_\uparrow}\bra{\psi_{\uparrow}}\right)\nonumber\\
    &+\frac{\Delta_c}{2} (\ket{\psi_\downarrow}\bra{\psi_{\uparrow}}+\text{H.c.}),
    \label{eq:Heff}
\end{align}
where $-M$ ($+M$) represents an effective magnetization of the state $\ket{\psi_\downarrow}$ ($\ket{\psi_\uparrow}$). 

\begin{figure}
    \centering    \includegraphics[width=0.5\textwidth]{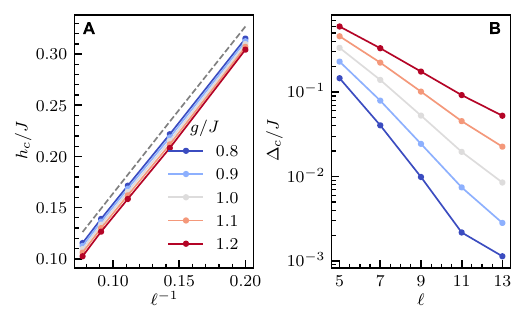}
    \caption{\textbf{Finite-size scaling of $\boldsymbol{h_c}$ and $\boldsymbol{\Delta_c}$ for $\boldsymbol{\beta=1.21}$.} (\textbf{A}) $h_c$ scales inversely with system size $\ell$ and is only weakly dependent on $g$. (\textbf{B}) The energy gap $\Delta_c$ becomes exponentially small with $\ell$, and increases with $g$.}
    \label{fig_finite_size}
\end{figure}

At each $\ell$ and $g$, we numerically calculate the ground-state and the first excited-state energy of the Hamiltonian in Eq.~(\ref{eq_H}) as a function of $h$ near $h_c$. We then fit the energy gap $\Delta(h)$ to the functional form $\Delta(h)=\Delta_c\sqrt{1+(h-h_c)^2/\delta^2}$ with fitting parameters $h_c$, $\Delta_c$ and $\delta$. This function matches the expression of the gap of the effective Hamiltonian $H_{\mathrm{eff}}$, with $\delta=\Delta_c/(2M)$. The result for $\beta=1.21$ is shown in Fig.~\ref{fig_finite_size}. When $g=0$, $h_c$ is given by $h_c =-(\sum_i\Delta h_i)/\ell\propto 1/\ell$, which corresponds to the gray dashed line in Fig.~\ref{fig_finite_size}A. At finite values of $g$, the value of $h_c$ only changes slightly from the $g=0$ case. $\Delta_c$ is a measure of the transition matrix element between $\ket{\psi_\downarrow}$ and $\ket{\psi_\uparrow}$, which requires all the spins to flip. The probability of the transition becomes exponentially small as the system size grows, but increases with the transverse field $g$. We empirically fit $\Delta_c$ to the functional form $\Delta_c\propto (g/g_*)^\ell$, where $g_*$ is a fit parameter and can be interpreted as the value at the paramagnetic to ferromagnetic phase boundary. For example, for $\ell=13$, $g_*=1.65 J$ for $\beta=1.21$, and $g_*=2.32 J$ for $\beta=0.78$. The familiar case of the Ising chain with nearest-neighbor interactions is recovered for $\beta\rightarrow \infty$, and therefore $\lim_{\beta\rightarrow\infty} g_* = J$.

\begin{figure}[t!]
    \centering   \includegraphics[width=0.5\textwidth]{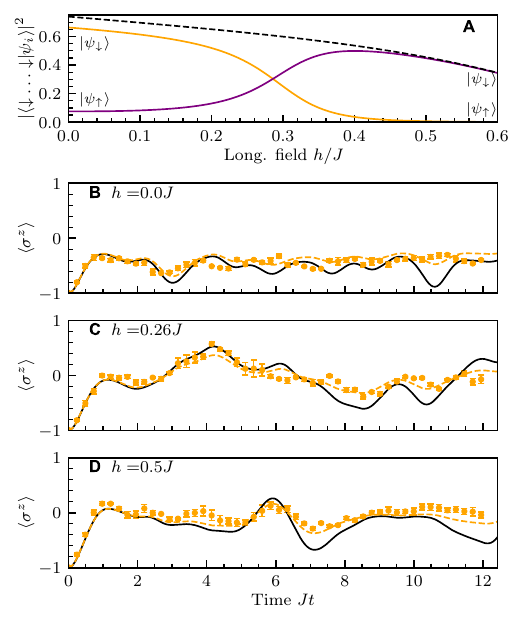}
\caption{\textbf{The $\boldsymbol{\ell=5}$ quench experiment.} 
 (\textbf{A}) Overlap between the initial state $\ket{\downarrow\dots\downarrow}$ and the two lowest-energy eigenstates $\ket{\psi_1}$ (orange line) and $\ket{\psi_2}$ (purple line) for the quench experiment in Fig.~\ref{fig_hc}B, for $\beta = 1.21$ and $g=1.2J$. The dashed line shows the sum of the two overlaps. The initial state has a large overlap with $\ket{\psi_\downarrow}$ away from $h_c$. However, the overlap with the two eigenstates becomes equal at $h_c$, resulting in large oscillations of $\langle\sigma_z\rangle$ after the quench. (\textbf{B}-\textbf{D}) Quench dynamics at $h=0$, $h=0.26J$, and $h=0.5J$, respectively. Points correspond to experimental data with 250 repetitions and the error bars indicate the standard deviation from $N=1000$ bootstrap samples. The maximum value from each time evolution is shown in Fig.~\ref{fig_hc}B. The solid black (dashed orange) line corresponds to numerical simulation without (with) decoherence.}
    \label{fig_quench_data}
\end{figure}

The effective Hamiltonian in Eq.~(\ref{eq:Heff}) provides an understanding of the results of the quench experiment in Fig.~\ref{fig_quench_data}.
For the time-evolved state $\ket{\psi(t)}=e^{-itH_{\mathrm{eff}}}\ket{\psi_\downarrow}$, the probability of measuring $\ket{\psi_\uparrow}$ at time $t$ reads
\begin{equation}
    P(\ket{\psi_\uparrow}, t) \coloneq |\langle\psi(t)|\psi_\uparrow\rangle|^2=\Big(\frac{\Delta_c}{\omega}\Big)^2\sin^2\Big(\frac{\omega t}{2}\Big),
\end{equation}
where
\begin{equation}
\omega^2=\Delta_c^2+4M^2(h-h_c)^2.    
\end{equation}
Therefore, for $M|h-h_c|\gg \Delta_c$, the probability of flipping to the state $\ket{\psi_\uparrow}$ is very small and the magnetization is approximately constant. On the other hand, for $h= h_c$, the system oscillates between $\ket{\psi_\downarrow}$ and $\ket{\psi_\uparrow}$, with a large oscillation amplitude in the magnetization. The time required to flip the magnetization diverges exponentially with the system size.

In our quench experiment, the initial state is the fully polarized state $\ket{\downarrow\dots\downarrow}$ rather than $\ket{\psi_\downarrow}$, so the oscillations are not expected to be perfect. Nevertheless, a peak of the oscillation amplitude for $h=h_c$ is still expected due to the large overlap of the initial state with $\ket{\psi_\downarrow}$ with the chosen parameters, as shown in Fig.~\ref{fig_quench_data}A. Examples of quench dynamics are shown in Fig.~\ref{fig_quench_data}B-D.

%%%
\subsection{13-spin eigenstates}
This section provides additional details of the 13-spin eigenstates $\ket{\psi_\uparrow}$ and $\ket{\psi_\downarrow}$ produced by the $g$-field ramp in Fig.~\ref{fig_hc}D.

\begin{figure}
    \centering    \includegraphics[width=0.35\textwidth]{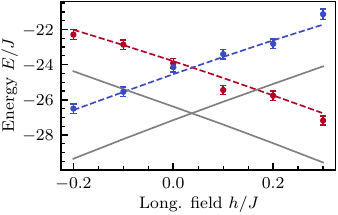}
    \caption{\textbf{Measured eigenenergies of the $\boldsymbol{\ell=13}$ system versus numerical simulation.} The red and blue points correspond to the measured energy of $\ket{\psi_\uparrow}$ and $\ket{\psi_\downarrow}$, respectively, at $\beta=0.78$, also shown in Fig.~\ref{fig_hc}D. The states are prepared via a $g$-ramp from $g=0$ to $1.7J$, as explained in the main text. The gray lines correspond to the numerical simulation of the state energies. The dashed lines correspond to the numerical simulation including a 2\% bit-flip error on each qubit, which fit the experimental points reasonably well.}
    \label{fig_13ion_energy}
\end{figure}

Figure~\ref{fig_13ion_energy} compares the measured energies of $\ket{\psi_\uparrow}$ and $\ket{\psi_\downarrow}$ for $\ell=13$ and $\beta=0.78$, also shown in Fig~\ref{fig_hc}D, to numerically computed energies of the states after the ramp (gray lines). To better match the experimental data, we account for a spin-flip error probability of $p=0.02$ per spin. We first compute the state $\ket{\psi}=U[H(t)]\ket{\psi_0}$, where $\ket{\psi_0}=\ket{\uparrow\dots\uparrow}$ or $\ket{\downarrow\dots\downarrow}$ and $U[H(t)]$ is the unitary evolution corresponding to a linear ramp of $g$ from $0$ to $1.7J$. We then generate the single-spin-flip error states $\ket{\psi}^{\alpha}_{i}=R^{\alpha}_i(\pi)\ket{\psi}$, where $\alpha=x$ or $z$, and $R^{\alpha}_i(\pi)$ is a $\pi$-rotation around the $\alpha$ axis (we choose $x$ and $z$ because the energy is measured in these basis). Next we calculate the energy including the error states $E_i=(1-2p)E(\psi)+p \sum_{\alpha}E(\psi^\alpha_i)$, and average over all spins $\overline{E}=\text{mean}(E_i)$, which is shown by the dashed lines in Fig.~\ref{fig_13ion_energy}. The bit-flip errors, which may occur in state-preparation and meansurement or during the Hamiltonian evolution, are more likely to increase the energy because $\ket{\psi_\uparrow}$ and $\ket{\psi_\downarrow}$ are the two lowest-energy eigenstates.

\begin{figure}
    \centering    \includegraphics[width=0.5\textwidth]{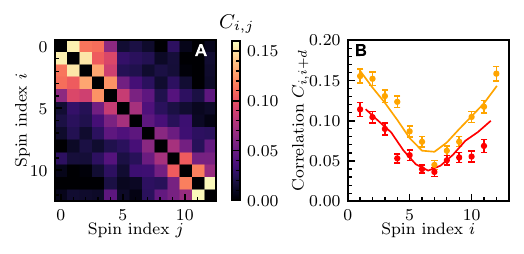}
    \caption{\textbf{Spatial correlations for the $\boldsymbol{\ell=13}$ eigenstate.}  (\textbf{A}) Connected correlation matrix $C_{i,j}=\langle\sigma_{i}^{z}\sigma_{j}^{z}\rangle-\langle\sigma_{i}^{z}\rangle\langle\sigma_{j}^{z}\rangle$ for $\ket{\psi_\downarrow}$ at $h=0.3J$ with $g=1.7J$ and $\beta=0.78$. (\textbf{B}) The nearest-neighbor (orange) and next-nearest-neighbor (red) $C_{i,i+d}$ ($d=1,2$) components from \textbf{A}, compared with numerical simulation (solid lines). Error bars represent standard deviation from $N=1000$ bootstrap samples. The correlation is stronger at the edges than the center due to the site-dependent longitudinal field $\Delta h_i$.
    }
    \label{fig_ground_state_corr}
\end{figure}

Figure~\ref{fig_ground_state_corr}A shows the connected two-point spin correlations $C_{i,j} = \langle \sigma^z_i \sigma^z_j \rangle - \langle \sigma^z_i \rangle \langle \sigma^z_j \rangle$ for the eigenstate $\ket{\psi_\downarrow}$ for $\ell=13$, $\beta=0.78$, and $h=0.3$ prepared with the $g$-ramp in Fig.~\ref{fig_hc}(D). These correlations are nontrivial compared to the product state $\ket{\downarrow\dots\downarrow}$ at $g=0$. Notably, nearest-neighbor correlations are stronger near the edges than at the center. This is due to the spatially varying longitudinal field $\Delta h_i$ that breaks translational symmetry, and enhances correlations near the boundaries, as evident from the nearest- and next-nearest-neighbor components of $C_{i,j}$ in Fig.~\ref{fig_ground_state_corr}(B).
\subsection{Optimized ramp}\label{sec_ramp}
This section describes the implementation of optimized ramp protocols designed to enhance adiabaticity across the phase transition while minimizing the total ramp duration.

To optimize the ramp through the phase boundary for the $\ell=5$ chain, we use a protocol that dynamically adjusts the ramp speed based on the instantaneous energy gap $\Delta(h)$. A simple linear ramp of the longitudinal field $h$ would require prohibitively long durations to maintain adiabaticity, as the energy gap $\Delta_c$ at the avoided crossing decreases exponentially with system size. Instead, our optimized ramp slows down near the minimum gap and speeds up elsewhere, allowing for near-adiabatic evolution in significantly shorter time.

\begin{figure}
    \centering
    \includegraphics[width=0.35\textwidth]{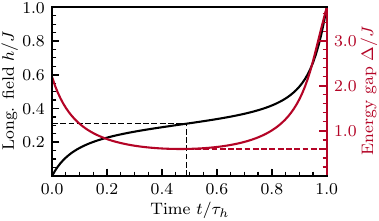}
    \caption{\textbf{Optimized $\boldsymbol{h}$-ramp for $\boldsymbol{\ell=5}$.} The black line shows the ramp trajectory $h(t/\tau_h)$, and the red line shows the instantaneous energy gap between the ground and the first excited states, for $\beta=1.21$ and $g=1.2J$. The dashed lines mark the critical field $h_c$, the corresponding critical time, and the energy gap $\Delta_c$.}
    \label{fig_ramp}
\end{figure}

The ramp is constructed such that the rate of change of the longitudinal field satisfies~\cite{Roland2002, Richerme2013}
\begin{equation}\label{eq_adiabatic}
\frac{dh}{dt}\propto \Delta(h)^2,
\end{equation}
where $\Delta(h)$ is the energy gap between the two lowest-energy eigenstates. The resulting ramp trajectory, obtained by numerically integrating Eq.~\eqref{eq_adiabatic}, is shown in Fig.~\ref{fig_ramp}. For comparison, a linear ramp with a slope equal to that at $h=h_c$ would require a total duration approximately 4.3 times longer.

We also applied optimized ramp profiles for the $g$-field ramps in Figs.~\ref{fig_ramp_5ions} and \ref{fig_ramp_13ions} to reduce total evolution time. Since the energy gap remains relatively large and varies smoothly during these ramps, the optimized profiles closely resemble linear ramps in practice.

\subsection{Nonadiabaticity of the optimized ramp protocol for $\ell=13$}\label{sec_13ion_optimized_ramp
}
This section presents experimental data demonstrating that for a 13-spin system, adiabatic passage through the first-order quantum phase transition is not achievable with experimentally feasible ramp protocols, due to the exponentially small energy gap near the transition. 

\begin{figure}[h]
    \centering    \includegraphics[width=0.35\textwidth]{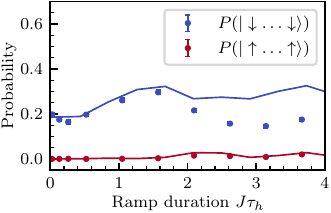}
    \caption{\textbf{
    State probabilities during the ramp for $\boldsymbol{\ell=13}$, $\boldsymbol{g=1.2J}$, and $\boldsymbol{\beta=1.21}$.} Points correspond to experimental data with 2000 repetitions and the error bars indicate the standard deviation from $N=1000$ bootstrap samples. The solid lines correspond to numerical simulation without decoherence while the points represent experimental data. To obtain these probabilities, the same ramp protocol as in Fig.~\ref{fig_ramp_5ions}A is used but for the larger system of $\ell=13$ spins. The state $\ket{\uparrow\dots\uparrow}$ remains unpopulated throughout the ramp due to the exponentially small $\Delta_c$. }   \label{fig_13ion_ramp_adiabatic}
\end{figure}

In Fig~\ref{fig_ramp_5ions}(A), we probe the adiabaticity of a Landau-Zener for an $\ell=5$ chain. Figure~\ref{fig_13ion_ramp_adiabatic} shows the result for the same ramp protocol for the $\ell=13$ chain. Despite the optimized ramp profile, no significant population transfer to $\ket{\psi_\uparrow}$ is observed, confirming that the exponentially small energy gap prohibits adiabatic evolution across the transition for this system sizes.

\subsection{Eigenstates for $h > h_c$}\label{sec_spectrum} 

This section presents a study of low-energy eigenstates in the regime $h > h_c$, providing insight into domain formation and the excitation pathways during the ramp. 

\begin{figure}[t!]
    \centering    \includegraphics[width=0.5\textwidth]{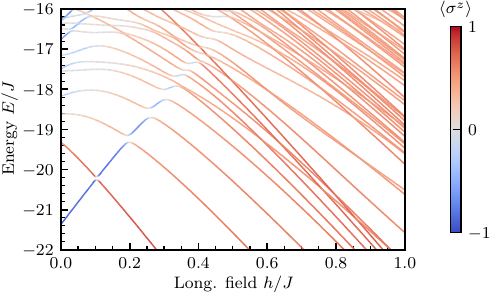}
    \caption{\textbf{Low-energy eigenenergies as a function of $h$ for $\boldsymbol{\ell=13}$, $\boldsymbol{g=1.2J}$, and $\boldsymbol{\beta=1.21}$.} Only the energy levels associated with eigenstates symmetric under left-right reflection are shown. Lines are colored according to the value of $\langle \sigma^z\rangle$. As $h$ increases, the eigenstate $\ket{\psi_\downarrow}$ forms many avoided crossings with other levels with fewer than $\ell$ total number of $\uparrow$-spins. These avoided crossings occur at several values of $h$, and the associated energy gaps can become wider than the critical gap $\Delta_c$, allowing these states with partial spin flips (positive magnetization) to become populated during the ramp.}
    \label{fig_N13Spectrum}
\end{figure}

The energy spectrum for $\ell=13$, $\beta=1.21$, and $g=1.2J$ is shown in Fig.~\ref{fig_N13Spectrum}. The Hamiltonian is invariant under left-right reflection, defined by exchanging spin indices $i \leftrightarrow \ell+1-i$ for all $1\leq i\leq \ell$. Consequently, its eigenstates can be classified as either symmetric or antisymmetric under this reflection. Because the initial state used in our experiment is symmetric, the system evolves entirely within the symmetric subspace (in the absence of experimental noise). For this reason, only symmetric eigenstates are shown in the energy spectrum.

To gain intuition about the spin polarization of the eigenstates, it is helpful to consider the limit $h \gg g$, where the longitudinal field dominates the Hamiltonian, and $\beta\gg 1$, where interactions are only nearest-neighbor. Then, the energy of an eigenstate is approximately determined by the number of $\uparrow$-spins, $N_\uparrow$, and the number of domain walls, $N_{\text{DW}}$. The dominant $h$-dependent contribution to the energy comes from the longitudinal-field term, yielding a leading-order slope $\sim 2(\ell - N_\uparrow)$ for the eigenenergy as a function of $h$. For a fixed $N_\uparrow$, states with fewer domain walls have lower energy due to minimized spin-spin interaction energy. In particular, the lowest-energy configuration for a given $N_\uparrow$ corresponds to all $\uparrow$-spins forming a single contiguous domain attached to either of the static $\uparrow$-spins at the chain edge.

For example, in Fig.~\ref{fig_N13Spectrum}, at $h/J \approx 0.55$ and $E/J=-22$, the eigenstate is adiabatically connected to the symmetric superposition $1/\sqrt{2}(\ket{\uparrow\dots\uparrow\downarrow}+\ket{\downarrow\dots\downarrow\uparrow})$ in the $h\gg g$ limit, where a single domain of 12 $\uparrow$-spins is connected to either edge. This state has an energy offset of $\sim 2h + 4J$ relative to $\ket{\psi_\downarrow}$, consistent with the expected linear-$h$ dependence and the interaction-energy shift.
More generally, for large $h$, the ordering of excited states depends on both $N_\uparrow$ and $N_{\text{DW}}$: reducing $N_\uparrow$ increases the energy due to the longitudinal field, and additional domain walls increases interaction energy.
	
In the limit $h\gg g$, the states most accessible from $\ket{\psi_{\downarrow}}$ are those containing only a few $\uparrow$-spins. During the $h$-ramp, these states become populated successively depending on the corresponding transition amplitudes and on the ramp speed. States with many $\uparrow$-spins have exponentially suppressed amplitudes, and thus are harder to reach, requiring very slow ramps. As $h$ increases, states with smaller $\uparrow$-spin domains come into resonance with $\ket{\psi_{\downarrow}}$; due to their larger transition amplitudes, they can be populated with moderate ramp speeds.

\subsection{Numerical simulation of the experiment}\label{sec_sim}
This section outlines the numerical methods used to simulate the experiment, including the modeling of coherent dynamics and decoherence effects.

Two methods are used to numerically simulate the time dynamics of the experiment. The first method is solving the Schr\"{o}dinger's equation
\begin{equation}
\frac{d}{dt}\ket{\psi(t)}=-i H(t)\ket{\psi(t)}.
\end{equation}
We use the $J_{i,j}$ values calculated using Eq.~\eqref{eq_Jij} below and normalize them so that the average of the nearest-neighbor interactions is equal to the experimentally measured value.

\begin{figure}[t!]
    \centering    \includegraphics[width=0.5\textwidth]{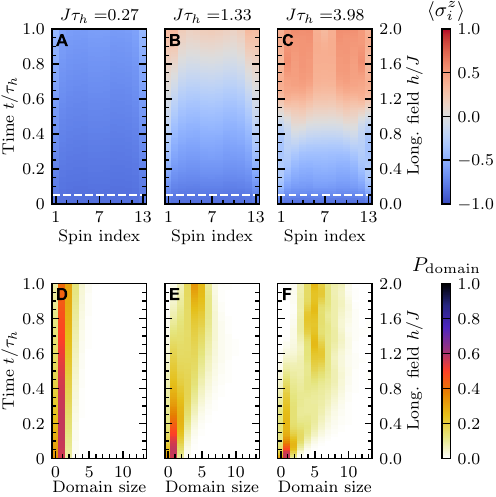}
    \caption{\textbf{Numerical simulation of linear-ramp dynamics for $\boldsymbol{\ell=13}$.} These data are numerically simulated counterparts of the data shown in Fig.~\ref{fig_ramp_13ions}.~(\textbf{A}-\textbf{C}) Time evolution of local magnetization $\langle \sigma_i^z \rangle$, and (\textbf{D}-\textbf{F}) time evolution of probability $P_\text{domain}$ of domains of given size for $J\tau_h=0.27$, $1.33$, and $3.98$, respectively.}
    \label{fig:13ionsim}
\end{figure}

The numerical simulation of local magnetization, $\langle \sigma_i^z \rangle$, and probability the largest flipped-spin domain to have size $n$, $P_\text{domain}(n)$, for the 13-ion linear-ramp experiment is shown in Fig.~\ref{fig:13ionsim}. These results should be compared with the experimental results in Fig.~\ref{fig_ramp_13ions}. 
The experimental data is in good agreement with the simulation. Many of the features from the simulation, such as the propagation of spin-flip domains from the edge to the bulk, and the formation of larger domains of flipped spins at longer ramp durations, are accurately reproduced by the experiment.

For the 5-ion ramp experiments, the total duration of the $g$- and $h$-ramps becomes long enough that decoherence causes a significant difference between the experiment and the exact dynamics. The experiment of this work has two main sources of decoherence. The first is a site-dependent dephasing in the Bloch $z$ basis with rate $\gamma_{i}$ for the $i$-th spin and the second is a common $x$-dephasing rate $\gamma_{c}$, see Methods, Sec.~\ref{sec_exp}. The open-quantum-system master equation including these two decoherence sources is given by~\cite{Feng2023}
\begin{align}
\frac{d\rho}{dt}=&-i[H, \rho]-\sum_{i}\gamma_{i}(\rho-\sigma^{z}_{i}\rho\sigma^{z}_{i})\nonumber\\
&-\gamma_{c}\sum_{i}(\rho-\sigma^{x}_{i}\rho\sigma^{x}_{i}),\label{eq:decoh}
\end{align}
where $\rho$ is system's density matrix and $S_{x}=\sum_{i}\sigma^x_i$. The values of $\gamma_{i}$ and $\gamma_{c}$ are obtained by fitting the decay of oscillation contrast of the $J_{i,j}$ measurement. For $\ell=5$, we get $\gamma_i=0.009J$ for all ions (since the 5 adjacent ions at the center of the chain have very similar decoherence rate), and $\gamma_c=0.010J$. For the $\ell=5$ ramp shown in Fig.~\ref{fig_ramp_5ions}, we see good agreement between the experiment and decoherence simulation, especially for longer ramp times. The $\ell=5$ quench dynamics, shown in Fig.~\ref{fig_quench_data}, also shows the decoherence effect at long evolution times. We do not simulate the 13-ion experiment with decoherence, because the linear ramps have shorter durations, and decoherence effects are less severe.

\subsection{String breaking and the boundary conditions}\label{sec_Z2}

This section presents details of our implementation of the boundary conditions that mimic static quarks. This set up yields a mapping between the mixed-field Ising model and the string-breaking problem in a gauge theory. 

\begin{figure}
    \centering    \includegraphics[width=0.5\textwidth]{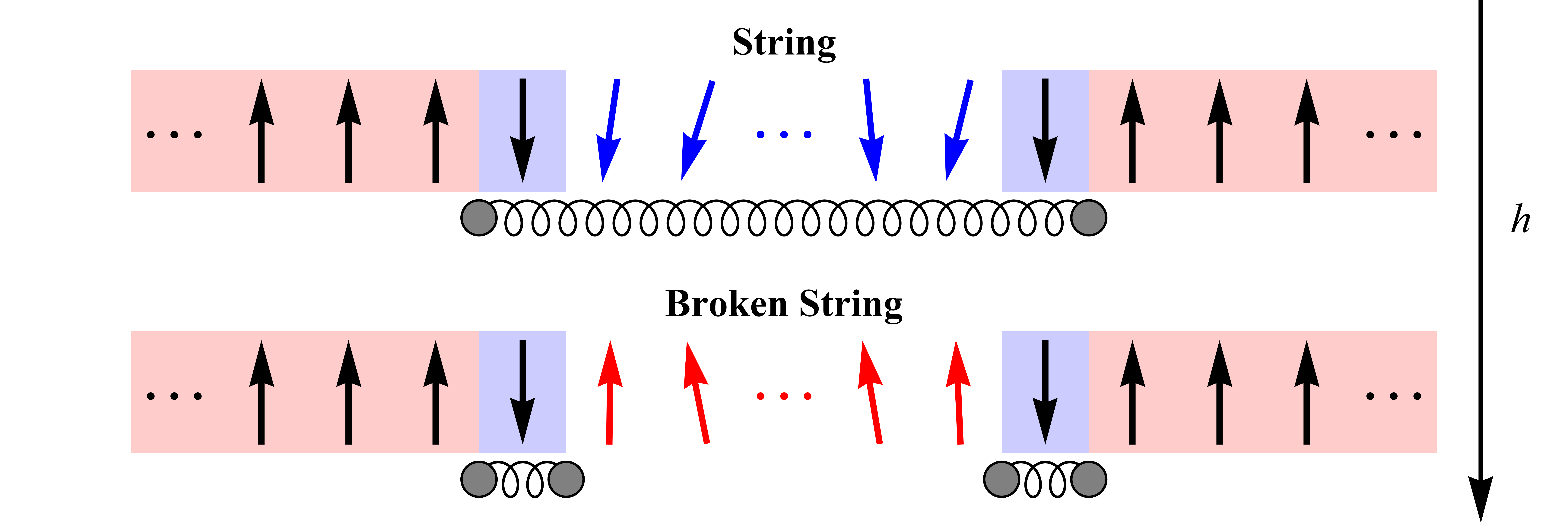}
    \caption{\textbf{Mapping of the spin model to a $\mathbf{Z}_2$ lattice gauge theory.} The gauge theory describes dynamics of quarks (gray circles) and electric fluxes or strings (springs connecting the quarks).
    }
    \label{fig_string_breaking}
\end{figure}

The mixed-field Ising model in Eq.~(\ref{eq_H}) can be mapped to a $\mathbb{Z}_2$ lattice gauge theory \cite{Lerose2020,Surace_2021,De2024}, as shown in Fig.~\ref{fig_string_breaking}. In this model, fermionic particles (quarks) are created or annihilated on the links between the spin sites, and the spins are mapped to the bosonic fields mediating the interaction between the fermions. A domain wall ($\uparrow\downarrow$ or $\downarrow\uparrow$) represents a quark. The $\downarrow$-spins in between two quarks represent the string joining them, whereas the $\uparrow$-spins represent the vacuum. The fermionic degrees of freedom can be integrated out by Gauss's law, and the dynamics can be entirely determined by the spin degrees of freedom.

In our model, the static spins on the boundaries represent two quarks at the edges, with the vacuum extending to infinity to the left (right) of the left (right) static quark. The static-spins' interactions with the dynamical spins create the site-dependent longitudinal field $\Delta h_i$ given by~\cite{Surace2024}
\begin{equation}
\Delta h_i=-\frac{1-2e^{-\beta}}{1-e^{-\beta}}\Big(e^{-\beta (i-1)}+e^{-\beta(\ell-i)}\Big).
\end{equation}
In our setup, where $\beta > \log(2)$, we find that $\Delta h_i$ is strictly negative. The amplitude of $\Delta h_i$ is largest at the two edges and decays exponentially towards the center.

For a finite chain of length $\ell$, the critical field satisfies  $h_c>0$, and the ground state just below the transition at $h=0<h_c$ is well approximated by $\ket{\psi_\downarrow}$, which 
represent the string state. The string potential energy can be adjusted by the longitudinal-field strength $h$, which imparts an energy cost on the $\downarrow$-spins as $h$ increases. At $h>h_c$, the ground state becomes $\ket{\psi_\uparrow}$, which represents the formation of two additional quarks at the edges, thus breaking the original string.

\subsection{Experimental Apparatus}\label{sec_exp}

This section describes our trapped-ion quantum simulator and the experimental techniques used to implement site-resolved spin interactions and programmable fields.

In the main text, we define the spin-spin interaction along the $z$-axis, following standard Ising model conventions. However, in our experimental implementation using the Mølmer-Sørensen interaction, the physical spin-spin coupling is realized along the $x$-axis. In this section, we adopt the original experimental convention: the qubit computational basis is the $z$ basis, the spin-spin interactions are along the $x$ axis, and field and measurement directions are described accordingly. To distinguish this basis, and their associated Pauli operators, from those in the main text, we adopt a tilde notation for states and operators in the following. 
To explicitly connect with the notation $\ket{\uparrow}$, $\ket{\uparrow}$, and $\sigma^{z
%/x
}_i$ used in the main text for states and the Pauli-$z$ operator, in the present appendix, one should consider the identifications $\ket{\uparrow} \equiv \ket{\tilde \uparrow}_x$, $\ket{\downarrow} \equiv \ket{\tilde \downarrow}_x$, and $\sigma^{z
%/x
}_i \equiv \tilde \sigma^{x
%/z
}_i$.

We trap a chain of $N=15$  $^{171}$Yb$^+$ ions in a surface ion trap \cite{Maunz2016}. The spin states are encoded in the two clock levels of the $^2S_{1/2}$
ground state, namely $\lvert\tilde\downarrow\rangle_{z} \coloneq \lvert F=0, m_{F}=0\rangle$ and $\lvert\tilde\uparrow\rangle_{z} \coloneq \lvert F=1,m_{F}=0\rangle$, with an energy difference $\omega_0=2\pi\times12.6$~GHz. At the begining of each experiment cycle, the ions are cooled with Doppler cooling, and the collective radial motional modes responsible for the spin-spin interactions are cooled by resolved-sideband cooling. The ions are initialized in the $\lvert\tilde\downarrow\rangle_{z}$ state by optical pumping, and prepared in either $\lvert\tilde\uparrow\rangle_{x}$ or $\lvert\tilde\downarrow\rangle_{x}$ with single-qubit $R_y(\pm \pi/2)$ rotations.  The experiments of this work involve performing the Hamiltonian ramps and measuring the state distribution in the end in the $x$ basis. This measurement amounts to applying $R_y(- \pi/2)$ rotations followed by state-dependent fluorescence detection.

Spin-spin interaction can be generated by virtually coupling to one set of the radial collective motional modes of the ion chain. We tune the orientations of the two radial principal axes of the trap, so that one principal axis is parallel and the other perpendicular to the $\vec{k}$-vector of the Raman beatnote. Thus, only one set of $N$ motional modes is affected by the Raman laser. For an $N=15$ ion chain, the collective motional modes range from the highest-frequency center-of-mass mode $\omega_{0}=2\pi\times3.03$~MHz to the lowest-frequency ``zig-zag'' mode $\omega_{N}=2\pi\times2.74$~MHz. 

The global and individual beams are derived from a pulsed 355~nm laser. The global Raman beam passes through an acousto-optic modulator (AOM), which splits the beam into two frequency components given by $\omega^{\text{global}}=\omega^{\text{global}}_{0}\pm(\omega_{N}+\mu)$, where $\mu$ is a large negative detuning from the zig-zag mode. The individual beam is split into more than 30 beams with a diffractive element, and each beam passes through a channel from a 32-channel AOM. Each individual AOM channel also contains two frequencies that results in the dual array of individual beams, which are imaged onto the ions through a telecentric optical path. The first array of individual beams has the frequency $\omega^{\text{ind}}_{0}$ that satisfies the condition $\omega^{\text{global}}_{0}-\omega^{\text{ind}}_{0}=\omega_{0}$. Thus the beatnote between $\omega^{\text{global}}$ and $\omega^{\text{ind}}_{0}$ corresponds to $\omega_0\pm(\omega_{N}+\mu)$. In the dispersive regime ($|\mu|\gg\eta\Omega_{i}$, where $\eta$ is the Lamb-Dicke parameter and $\Omega_{i}$ is the Rabi frequency on the $i$-th ion), the simultaneous red and blue sidebands in the Raman beatnote produce the spin-spin interaction term $\sum_{i<j} J_{i,j}\tilde\sigma^{x}_{i} \tilde\sigma^{x}_{j}$. The interaction strength is given by
\begin{equation}\label{eq_Jij}
J_{i,j}=\sum_{k=1}^{N}\frac{\eta^{2}b_{i,k}b_{j,k}\Omega_{i}\Omega_{j}}{2(\omega_{N}+\mu-\omega_{k})},
\end{equation}
where the summation is over all the participating motional modes, $\eta=0.08$ is the Lamb-Dicke factor, and $b_{i,k}$ is the mode-participation matrix element of the $i$-th ion in the $k$-th motional mode. In this work, the negative detuning $\mu$ means that the interaction is dominated by the zig-zag mode. The interaction matrix $J_{i,j}$ decays exponentially with ion separation, $J_{i,j}\approx J\exp[-\beta(|j-i|-1)]$, where $\beta=1.21$ for $\mu=-2\pi\times100$~kHz, and $\beta=0.78$ for $\mu=-2\pi\times35$~kHz. 
\begin{figure}[t!]
    \centering   \includegraphics[width=0.475\textwidth]{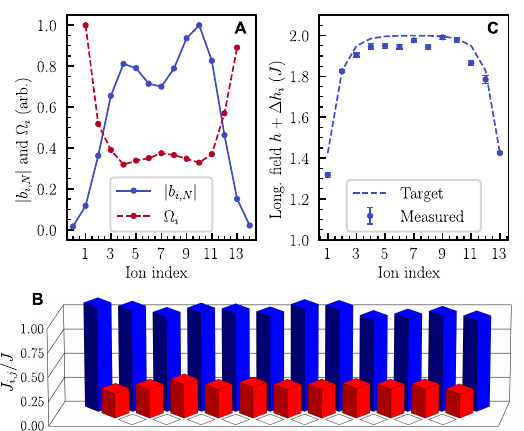}
    \caption{\textbf{Calibration of the Hamiltonian terms.} (\textbf{A}) Blue points correspond to the measured $|b_{i,N}|$ values for the zig-zag mode, normalized by the maximum value. Red points correspond to $\Omega_i$ in Eq.~(\ref{eq_Jij}) to make a uniform $J_{i,j}$ matrix. (\textbf{B}) Measured nearest-neighbor (blue) and next-nearest-neighbor (red) components of the $J_{i,j}$ matrix, normalized by $J$, for $\ell=13$ and $\beta=1.21$. (\textbf{C}) Measured site-dependent longitudinal-field values vs. target values, for $\ell=13$, $\beta=1.21$, and $h=2J$.}
    \label{fig_Jij}
\end{figure}

For the zig-zag mode, the mode-participation amplitude $|b_{i,N}|$ is higher at the center of the chain and lower at the edges. To compensate for this inhomogeneity and generate a translationally invariant $J_{i,j}$ matrix, we use RF amplitude control for the individual AOM channels to tune the local Rabi frequencies $\Omega_i$ so that the nearest-neighbor interactions are uniform, as demonstrated in Fig.~\ref{fig_Jij}A. These RF amplitudes (in arbitrary units) are used as the $\Omega_i$ values in Eq.~\eqref{eq_Jij} to compute the $J_{i,j}$ matrix numerically. We then rescale the resulting matrix so that the average nearest-neighbor interaction matches the experimentally measured value.

We experimentally measure $J_{i,j}$ by turning on only the individual beams at the $i$-th and $j$-th ions and observing the Rabi oscillation between $\ket{\tilde\uparrow_i\tilde\downarrow_j}z$ and $\ket{\tilde\downarrow_i\tilde\uparrow_j}z$. The measured nearest-neighbor and next-nearest-neighbor $J_{i,j}$ components for $\beta=1.21$ are shown in Fig.~\ref{fig_Jij}B. In addition, the sign of $b_{i,N}$ is staggered, i.e., $\text{sgn}(b_{i,N})=-\text{sgn}(b_{i+1,N})$, which results in the opposite sign between the $d$-th nearest-neighbor and $d+1$-th nearest-neighbor interaction, i.e., $\text{sgn}(J_{i,i+d}) = -\text{sgn}(J_{i,i+d+1})$. We add a $\pi$ phase shift to every other individual beam to render all the interaction signs the same.

In the experiment, we always maintain an $N=15$ ion chain. The voltages on the trap electrodes are adjusted so that the center 13 ions are aligned with the evenly-spaced individual beams.  We use the AOMs to turn on the individual beams for the center $\ell$ ions to simulate the different system sizes. For $\ell=5$, $J=2\pi\times 0.68$~kHz, and for $\ell=13$, $J=2\pi\times0.4$~kHz.

The second array of individual-addressing beams have the frequency $\omega^{\text{ind}}_{1}=\omega^{\text{ind}}_{0}-(\omega_{N}+\mu)$. The beatnote between $\omega^{\text{ind}}_{1}$ and the lower-frequency global tone is resonant with the qubit transition $\omega_{0}$. This produces a $\tilde\sigma^{\phi}_i=\cos\phi \, \tilde\sigma^x_i+\sin\phi \, \tilde\sigma^y_i$ rotation where $\phi$ is tuned by the phase of the RF drive of the individual AOM channel. We tune the phases of the individual RF tones to ensure $\tilde\sigma^{\phi}_i=\tilde\sigma^x_i$, and the amplitudes to ensure the measured $h+\Delta h_i$ frequencies match the desired values, as shown in Fig.~\ref{fig_Jij}C. The RF amplitudes are also varied in time to simulate the $h$-ramps.

Lastly, the time-dependent transverse field $g(t) \, \tilde\sigma^z_i$ can be produced by shifting the frequencies of both arrays of the individual beams by $2g(t)$. The varying intensity of the individual beams across the ion chain also produces site-dependent AC Stark shifts on the ions, which causes a coherent $\delta_i\tilde\sigma^z_i$ error. We add an equal and opposite frequency shift $-\delta_i$ to both individual RF tones to cancel the error.

Decoherence in our system primarily arises from fluctuations in the Rabi rates and motional-mode frequencies. Uncertainty in the relative positions between the ions and the tightly focused individual-addressing beams, caused by axial motional heating and trap charging, leads to site-dependent fluctuations in the Rabi frequency and results in dephasing along the Bloch $x$ axis, characterized by a rate $\gamma_i$. Additionally, fluctuations in the radial mode frequencies, arising from radial motional heating or variations in the trapping potential, lead to collective dephasing along the Bloch $z$ axis, with rate $\gamma_c$. Since the spin-spin interaction terms are driven by a much stronger Rabi rate ($\sim 2\pi \times 100$ kHz) compared to the longitudinal-field terms ($\sim 2\pi \times 1$ kHz), the decoherence rates $\gamma_i$ and $\gamma_c$ are determined from the measured contrast decay of the $J_{i,j}$-driven spin-spin oscillations.

\begin{acknowledgments}
This material is based upon work supported by the U.S.~Department of Energy (DOE), Office of Science, National Quantum Information Science Research Centers, Quantum Systems Accelerator. Additional support is acknowledged from the following agencies. This work was supported in part by the National Science Foundation's (NSF's) Quantum Leap Challenge Institute for Robust Quantum Simulation (award no.~OMA-2120757). F.M.S.~acknowledges support provided by the U.S.~Department of Energy (DOE) QuantISED program through the theory consortium ``Intersections of QIS and Theoretical Particle Physics'' at Fermilab, and by Amazon Web Services, AWS Quantum Program. A.L.~acknowledges funding through a Leverhulme-Peierls Fellowship at the University of Oxford. B.W., A.S., and A.V.G.~were also supported in part by AFOSR MURI, DOE ASCR Quantum Testbed Pathfinder program (awards No.~DE-SC0019040 and No.~DE-SC0024220), NSF STAQ program, DARPA SAVaNT ADVENT, ARL (W911NF-24-2-0107), and NQVL:QSTD:Pilot:FTL. B.W., A.S., and A.V.G.~also acknowledge support from the U.S.~Department of Energy, Office of Science, Accelerated Research in Quantum Computing, Fundamental Algorithmic Research toward Quantum Utility (FAR-Qu).
Z.D.~further acknowledges support by the DOE, Office of Science, Early Career Award (award no.~DE-SC0020271); as well as by the Department of Physics; Maryland Center for Fundamental Physics; and College of Computer, Mathematical, and Natural Sciences at the University of Maryland, College Park. Z.D. is further grateful for the hospitality of the Excellence Cluster ORIGINS at the Technical University of Munich, where part of this work was carried out. The research at ORIGINS is supported by the Deutsche Forschungsgemeinschaft (DFG, German Research Foundation) under Germany's Excellence Strategy (EXC-2094–-390783311).
\end{acknowledgments}

\newpage

\end{document}